\documentclass[graybox]{svmult}

\usepackage{type1cm}        
\usepackage{makeidx}         
\usepackage{graphicx}        
\usepackage{multicol}        
\usepackage[bottom]{footmisc}

\usepackage{newtxtext}       %
\usepackage{newtxmath}       

\usepackage[utf8]{inputenc}
\usepackage{color}
\usepackage[hyphens]{url}
\usepackage[breaklinks]{hyperref}

\usepackage{soul}

\makeindex             

\begin{document}

\title*{Mobile Phone Data for Children on the Move: Challenges and Opportunities}

\author{Vedran Sekara*, Elisa Omodei*, Laura Healy, Jan Beise, Claus Hansen, Danzhen You, Saskia Blume and Manuel Garcia-Herranz}
\authorrunning{Sekara et al.}
\institute{Vedran Sekara\at UNICEF Office of Innovation \email{vsekara@unicef.org}
\and Elisa Omodei \at  UNICEF Office of Innovation \email{eomodei@unicef.org}
\and Laura Healy \at UNICEF Data Research and Policy \email{lhealy@unicef.org}
\and Saskia Blume \at UNICEF Data Research and Policy \email{sblume@unicef.org}
\and Jan Beise \at UNICEF Data Research and Policy \email{jbeise@unicef.org}
\and Claus Hansen \at UNICEF Data Research and Policy \email{chansen@unicef.org}
\and Danzhen You \at UNICEF Data Research and Policy \email{dyou@unicef.org}
\and Manuel Garcia-Herranz \at UNICEF Office of Innovation \email{mgarciaherranz@unicef.org}
\and *These authors contributed equally to this work.}

%
%
\maketitle
\vspace{-8em}
\abstract{Today, 95\% of the global population has 2G mobile phone coverage~\cite{gsma2017} and the number of individuals who own a mobile phone is at an all time high. Mobile phones generate rich data on billions of people across different societal contexts and have in the last decade helped redefine how we do research and build tools to understand society.
As such, mobile phone data has the potential to revolutionize how we tackle humanitarian problems~\cite{datarevolution}, such as the many suffered by refugees all over the world. While promising, mobile phone data and the new computational approaches bring both opportunities and challenges~\cite{blumenstock2018}. Mobile phone traces contain detailed information regarding people's whereabouts, social life, and even financial standing.
Therefore, developing and adopting strategies that open data up to the wider humanitarian and international development community for analysis and research while simultaneously protecting the privacy of individuals is of paramount importance~\cite{undg}.
Here we outline the challenging situation of children on the move and actions UNICEF is pushing in helping displaced children and youth globally, and discuss opportunities where mobile phone data can be used. 
We identify three key challenges: data access, data and algorithmic bias, and operationalization of research, which need to be addressed if mobile phone data is to be successfully applied in humanitarian contexts.
}

\section{Introduction: Children on the move}
Millions of children are on the move across international borders fleeing violence, conflict, disaster, poverty, or in pursuit of a better life. As of 2017, 30 million are living in forced displacement, including 12 million child refugees and child asylum seekers, and 17 million children living in internal displacement due to conflict and violence~\cite{unhcr2018_2,idmc2}.
The conflict in the Syrian Arab Republic alone is estimated to have displaced 5.6 million people.
Of these, around 3.6 million are in Turkey, 950,000 in Lebanon, and 675,000 in Jordan~\cite{unhcr2018}.

Serious gaps in the laws, policies and services which are meant to protect children on the move further limit their access to protection and care.
When world leaders adopted the Global Compact on Refugees and the Global Compact for Migration in December 2018~\cite{refugee_compact,migration_compact}, they acknowledged the urgent and unmet needs of vulnerable child migrants and refugees. These Compacts were negotiated and agreed against a political backdrop where increasing numbers of people are leaving their countries of origin, due to an interplay of complex factors including conflict, economic circumstances, and a changing climate. 
The agreement of the Global Compacts was a great achievement, yet they have coverage gaps.
Concretely, the 40.3 million people who are internally displaced by armed conflict and generalized violence, are not protected under either of the two Global Compacts~\cite{idmc}. 
Irrespective of definition, refugees, migrants and internally displaced children share similar vulnerabilities and needs. 

As states struggle to manage migration and refugee flows, children are often at risk of being left in conditions that would be deemed unacceptable for native-born children, ending up in overcrowded shelters or makeshift camps. Too many still end up in immigration detention, despite recognition of the long-lasting and devastating impact on a child's development~\cite{keller2003impact}. Further, mistrust of authorities and fear of detention and deportation keep children from coming forward to seek protection, access to essential services and support, instead they chose a life on the streets. 

Along their route and at their final destination, refugees and migrants are further exposed to a plethora of other issues including discrimination, segregation, abuse or xenophobia~\cite{unicef-2017} -- factors which traditionally have proved to be hard to accurately monitor, understand and mitigate. 

In order to properly protect migrants and refugees there is a need to strengthen the evidence base, deliver humanitarian assistance at local, national and regional levels, and to bridge data gaps allowing States, non-governmental organizations (NGO), humanitarian agencies and the development sector to get a real sense and scale of the issue.

\subsection*{Data gaps}

Reliable, timely and accessible data and evidence are essential for understanding how migration and forcible displacement affect children and their families -- and for putting in place policies and programmes to meet their needs. Despite greater efforts over the past decade, we still do not know enough about children on the move: their age and sex; where they come from, where they are going and why they move; whether they move with their families or alone, how they fare along the way, what their vulnerabilities are, what they need, and how migration and asylum policies affect them~\cite{IOM}.

Data gaps make it difficult to get a real sense of the scale and patterns of global migration. In many cases data are not regularly collected, and quality is often poor. These problems are many times worse when it comes to data on migrant and forcibly displaced children, given the even greater challenges of measurement. Information comes from a patchwork of sources that provide little comparable global or even regional-level data. 

In addition, variations in the laws, definitions, rights and entitlements that apply to children further hamper comparisons between countries. Data are even scarcer on children moving undocumented across borders, those displaced, stateless or migrating internally, children left  behind by migrant parents, and those who have gone missing or lost their lives during dangerous journeys.

To ensure the protection of the rights of migrant and refugee children, national authorities, regional bodies and development partners can benefit from tools and instruments providing real time data on children on the move as well as better insights and understanding of the causes and consequences of the issues they face. Mobile phone data is uniquely positioned to answer some of these issues, especially given that mobile phones permeate every strata of society more prominently than any other communication technology generating high volumes of data on a daily basis~\cite{ungp}.


\section{Mobile data: Challenges and opportunities}
Mobile phone data has been used to map populations and their changes over time~\cite{deville2014dynamic}, understand human mobility patterns~\cite{gonzalez2008understanding}, and has been applied to validate theoretical models which traditionally are used to estimate movements patterns when no data is available~\cite{simini2012universal,barbosa2018human}.
Although mobile phone data mainly represents adult populations, as children are less likely to own a mobile phone, it can nevertheless be used in combination with other data sources (e.g. surveys) to understand youth mobility~\cite{christensen2011children}. 

In terms of applications, previous work applied mobile phone data to estimate population displacements after natural disasters~\cite{lu2012predictability} and to understand collective behavior during emergencies~\cite{bagrow2011collective}, demonstrating that people's reactions to exogenous events can be quantified and even predicted. Others applied human mobility estimated from mobile phone data to predict the geographic spread and timing of an epidemic~\cite{wesolowski2015impact}. As disease go where people travel, this methodology can be used to generate fine-scale dynamic risk maps~\cite{wesolowski2012quantifying}.

Beyond short-term mobility estimations, mobile phone data has also been used to estimate internal migration, for instance temporary and circular migration in Rwanda~\cite{blumenstock2012inferring} and climate change driven migration in Bangladesh~\cite{lu2016unveiling}. SIM cards are however linked to national providers and human mobility calculated from phone records can hence be used only to estimate internal displacements, as individuals usually get a new SIM card when they move to different countries, and even if they do not, mobile phone operators usually only record such events to have taken place outside of the country, but not the specific location. Hence, to study international migration patterns, alternative sources of data have been used, such as geo-tagged tweets~\cite{zagheni2014inferring} and Facebook data~\cite{zagheni2017leveraging}. Twitter data has also been used to estimate the relationship between short-term mobility and long-term migration~\cite{fiorio2017using}. 



In principle, mobile phone data coupled with tools from network science, algorithms from machine learning and artificial intelligence techniques have the potential to allow humanitarian and international organizations to address key issues and uncover novel insights. This includes a plethora of applications from: mapping socioeconomic vulnerabilities~\cite{blumenstock2015predicting,eagle2010network}, tracking epidemics in real-time~\cite{mcgowan2019collaborative}, to establishing causal relationships between factors such as climate change and migration.
This “data revolution“ promises to transform the international development and humanitarian sectors~\cite{datarevolution}. However, this data and its use in this context comes with its own type of challenges. 

Below we highlight three issues we find particularly important and which need to be addressed for mobile phone data to be used for the benefit of the most vulnerable. 

\subsection*{Data access: quality, privacy and usefulness}
Mobile phone data is originally not collected for scientific purposes, but for other reasons such as billing, as companies need to know who, when and how long people call each other to be able to correctly bill their customers. In addition, this data is highly sensitive and contains detailed information regarding people's social graphs, whereabouts, financial standing, and countless other behavioral patterns~\cite{blondel2015survey}. 
Consequently, national telephone operators and regulators are understandably cautious to share data with third parties including researchers and international organizations. 

Certain data initiatives such as the \textit{Data for Refugees}~\cite{salah2018data} (D4R) and the two previous \textit{Data for Development}~\cite{blondel2012data,de2014d4d} (D4D) challenges have had great success in opening up data to the broader scientific community. In part, this success is due to strong collaborations with telephone operators and with privacy researchers. Nonetheless such initiatives are rare.

Historically, access to mobile phone data has mainly occurred through one-to-one agreements between telephone operators and academic institution, international agencies or humanitarian organizations through non-disclosure or other legal agreements. Data sharing is further complicated by the lack of data anonymization and aggregation standards. 
Four different privacy-conscientious models have been proposed that balance privacy and usefulness of data~\cite{de2018privacy}. This includes \textit{limited release} models, similar in type to the D4R and D4D challenges, where a limited data sample (in terms of people and time) is shared with a small group of trusted affiliates.
Another framework is to give researchers and practitioners from humanitarian and international sectors remote access to anonymized data on a virtual environment controlled by the mobile phone operator. While the \textit{remote access} model is more secure it requires mobile phone operators to invest in infrastructure and technical expertise. \textit{Question and answer} frameworks have also been suggested. In this model, data stays within the premises of mobile phone operators and researchers can interact with it by submitting code (i.e. questions) to the system, which takes the code, validates and runs it, and returns results through an application interface (API). While the privacy benefits of this model are great, the approach requires substantial investments in: infrastructure, methods to validate submitted code, and developing systems that prevent leakage of personally identifiable information. The last approach is the \textit{aggregated data} model. Here the privacy-utility trade-off is balanced by only sharing indicators that are sufficiently disassociated from individual behavior. Examples include sharing pre-computed high level indicators such as radius of gyration (a measure of the average distance travelled by individuals) or social diversity (a measure of entropy), which are harder to link back to individuals.

There is unfortunately no one-fits-all data sharing model.
We have found the aggregated data model to work well in situations where there already exists a great body of literature, such as in using human mobility for epidemic modeling~\cite{wesolowski2012quantifying,bengtsson2015using,wesolowski2015impact}, given proper aggregation standards. 

For applications where standardized data aggregation frameworks have not been agreed upon, such as detecting causal relationships, the other data sharing models are exceedingly more useful.
In these situations initiatives like D4R are essential to develop and showcase new methodologies. 
Unfortunately, such initiatives are rare and limited to single countries. In humanitarian and development contexts there is a need for having access to data on a more robust basis. 
For example, if we are to achieve the sustainable development goals and eradicate poverty by 2030~\cite{sdgs} being able to accurately predict poverty is a good first step~\cite{blumenstock2015predicting}, yet if we are to succeed we need to be able to efficiently monitor progress aimed at combating poverty. This can only be done with continuous access to data. 


\subsection*{Data representativeness and bias}
A second key issue of using mobile phone records for humanitarian purposes is the question of how representative the data is of the overall population -- and of the most vulnerable groups in particular. It is in fact precisely the most vulnerable populations, and namely children, which tend to be the least represented in these newly available datasets built on the basis of technology usage.

In Turkey, for example, mobile subscriber penetration is 65\%~\cite{gsmamena2018}, meaning that 35\% of the Turkish population does not have a sim card registered under their name. On the other hand, some individuals own more than one SIM card. Partially, this is because some people make use of multiple phone numbers (personal, work, etc.), but in several cases this is instead due to the fact that people register multiple SIM cards under one name, for example, to be used by family members. In this sense, the Global System for Mobile Communications Association (GSMA) reported that in 2016 subscriber penetration in Turkey was only 43\%, but the connection penetration (i.e. number of unique phone numbers) was as high as 89\%~\cite{gsmamena2016}.
The issue is not that the data does not cover 100\% of the population (this is never the case in social science research, where surveys are always performed on a selected sample of the population, which also carries its limitations~\cite{carr2013missing}), but rather that the population captured in the data is not guaranteed to be representative of the entire population, especially the most vulnerable. Lower income individuals are for example less likely to own a phone than richer individuals, which results in their underrepresentation in datasets built by randomly sampling users. A study has shown, for example, that women travelling with children in sub-Saharan African countries are less likely to own phones (and, if they do, they are also less likely to use them more than once per day) than general travellers~\cite{marshall2016key}. Hence, datasets provided by mobile phone companies should be built by accurately selecting representative demographics among their clients, which can be done using the demographic information provided by users when subscribing, or based on phone usage patterns. Otherwise, the insights and findings obtained from biased datasets might not be accurately describe the dynamics of the most vulnerable.

Even after selecting a representative subscribers subset, additional sources of bias also need to be taken into account.
CDRs provide information on users only when a call is made/received or when a text is sent/received. Hence, calling/texting frequency plays a role in how much information on the user's behavior (e.g. their location over time) can be obtained. 
Therefore, individuals with a limited calling activity, who are normally the poorer ones, generate less data~\cite{leo2016socioeconomic}. This calls for special attention in the way we define algorithms. For example, the time window selected to compute mobility is a critical factor for bias~\cite{saramaki2015seconds}. If too short, it can enhance the bias of the data against the poorest individuals and vulnerable groups such as, for example, the women travelling with children from the study mentioned above.
Nowadays, providers can also record phone activity that goes beyond calling and texting, at least for smartphone users, whose internet usage can be recorded too (commonly known as XDRs). This reduces the sampling bias linked to calling/texting frequency, but exacerbates socioeconomic bias, as owning a smartphone and using Internet data is correlated with higher income.

Mobile network coverage is another important source of bias. In predominantly rural areas where tower density is lower, the spatial resolution of the information provided by CDRs is significantly more coarse grained, since each tower has a wider geographical coverage and hence the recorded user's location is less precise. Moreover, network absence in some areas will also limit user's behavior linked to mobile phone usage.

A notable study comparing socioeconomic surveys with mobile phone data representing daily movements of about 15 million individuals in Kenya showed that mobility estimates obtained from mobile phone records are surprisingly robust to biases in phone ownership across different geographical and socioeconomic groups~\cite{wesolowski2013impact}. Yet this is not the case for all regions and countries. A similar study for Rwanda found that phone owners are considerably wealthier, more educated and predominantly male~\cite{blumenstock2012divided}. Further testing across different countries and vulnerable groups is still needed. 

Much attention also needs to be used when building datasets and when interpreting and generalizing results obtained from these data. Models should not be blindly applied to a different context than the one they were originally developed for and tested in. Most computational social sciences studies have in fact been carried out largely using data from developed countries and are thus not highly representative of the poorest trenches of the population~\cite{blumenstock2018estimating}. This caution applies to spatial/geographical settings (i.e. using a model trained and tested in one specific country to make estimations or forecasts in another country) but also to temporal ones (i.e. a model trained and tested on data produced in a specific time window should be used with caution to make estimations or forecasts for future months/years)~\cite{lazer2014parable}.

\subsection*{Operationalizing research}
Data challenges like D4R and D4D -- in which private sector companies share a curated dataset with the research community to boost the development of new insights and methodologies on societal issues -- are notable initiatives encouraging scientific endeavour for social good. However, as mentioned earlier, in order for this data to be used in a consistent way for humanitarian purposes by international organizations like UNICEF, these efforts need to become systematic and integrated into existing frameworks.
Most of these scientific advances currently end up only living in scientific publications (which is a necessary first step to advance knowledge and guarantee the quality of the research) and, in some cases, in open repositories such as \textit{Github}. Some researchers also make an additional effort and build software that lives on dedicated websites or applications. However, these are usually isolated platforms that are not integrated with existing systems that governments and international agencies use in their daily operations. An example of effort towards this integration is UNICEF's \textit{MagicBox}~\cite{magicbox}, an open-source software platform that enables collaboration and the use of new data sources and computational techniques, like artificial intelligence and machine learning, for good.

The availability of one-off historical datasets is fundamental to make scientific advances, from discovering human behavior patterns to training and testing mathematical and computational models. In this sense data for research should be anonymized but as disaggregated as possible in order to allow data scientists and modellers to gain meaningful insights.

On the other hand, once models and data analysis pipelines have been finalized, their operationalization requires data streams which can be aggregated but need to be updated in near real-time. 
Furthermore, models need to integrate with data that comes from existing systems taking into account real-time changes of the situation, such as interventions like vaccine delivery, information campaigns, etc.

Hence, models running on real-time data should also learn in real-time, using techniques such as data assimilation~\cite{shaman2012forecasting}.

In order for all these efforts to be sustainable and work in the long term, robust ecosystems need to be built for collaboration. This means the creation of pipelines to allow joint research to be conducted with a strong focus on the most vulnerable, data explorations and models packaged into open-source modules to be reused and adapted to different contexts; and implementations that easily integrate with the existing systems already in place.

\section{Perspectives}
Beyond the aforementioned challenges, there is an additional and more fundamental challenge; a general disconnect between the scientific communities that work with "Big Data" and the humanitarian and development sector. This gap tends to produce oversimplifications and limited views of what is possible or needed from one community to the other. 
Data science is a complicated field with non-trivial possibilities and challenges.
Migration and forced displacement are complex problems with many causes, consequences and points of action.

In this complex ecosystem, UNICEF has identified six action points to keep every child uprooted by war, violence and poverty safe~\cite{6-point-agenda}. We believe the academic community can play a vital role in addressing these issues by working closely together with the humanitarian sector. Below we outline each action point tying them in with existing scientific literature within computational social science and mobile phone research. We hope this serves as an starting point for further dialogue to expand areas of collaboration.

\begin{itemize}
\item \textbf{Press for action on the causes that uproot children from their homes.} 
Properly understanding and monitoring the causes that lead to displacement is key to transition from reactive to proactive strategies. This is an exciting area of opportunity where novel research and new sources of data can play a mayor role to radically transform crisis response into crisis avoidance. Significant work has already been conducted on using different data sources from mobile phones to self reported data to monitor the fingerprint of some of the known drivers of migration~\cite{toole2015tracking,obradovich2017climate}. 
To radically change the way we face the growing problem of forced displacement additional research on causal relationships, tipping points, and monitoring strategies is critical.
\\
\item \textbf{Help uprooted children to stay in school and stay healthy.} 
Access to education, health care and other essential services is vital for children to succeed and have a good life. As such, quantifying and mapping gaps in basic services is critical for designing interventions.
Studies have shown that data collected by mobile phone operators can provide accurate and detailed population maps in privacy preserving ways~\cite{deville2014dynamic}.
Future avenues of research could look into adapting these methods to detect underserved populations, thus making it possible to design intervention schemes centered around health and education, such as determining where new schools and hospitals should be placed. 
Further, smartphones have the potential to track well-being and mental health in greater detail and at individual level~\cite{madan2012sensing,stopczynski2014measuring}.
\\
\item \textbf{Keep refugee families together.} 
Children who are separated from their families are more vulnerable to violence and abuse.
Recent studies have looked at how the strength of social ties decays across time and distance~\cite{raeder2011predictors,park2018strength}. Extending this work into also looking at the effects of family separation and forced displacement can provide critical insights for advocacy and mitigation strategies. 
In addition, phone data has been used to analyze the relationship between mobility patterns and social ties~\cite{toole2015coupling}, opening the door to identify potential friends and family based on historical phone registries. Despite the obvious privacy and ethical challenges, this opens the door to think about solutions to bring together families and communities that have been lost apart. 
It is, nevertheless, not yet clear how these effects scale to displaced populations or even different cultures. This illustrates that the current body of scientific literature has mainly focused on data from non-vulnerable populations and might therefore not be applicable to those most in need.
\\
\item \textbf{End the detention of refugee and migrant children by creating practical alternatives.} 
Detention is harmful to children's health and well-being and undermines their development. 
Mobile phone data has been applied to categorize social networks~\cite{eagle2009inferring}, identify communities~\cite{expert2011uncovering} and understand urban environments in terms of social dynamics and segregation~\cite{louail2014mobile}. These findings put together can help in better identifying and advocating for alternatives to detention, helping identify places and communities specially well suited to host certain displaced children.
In addition, is is of paramount importance to better understand and quantify the horrible effects that detention has on children. Research has shown that social signatures can be quantified over time~\cite{saramaki2014persistence}, opening up possibilities to measure the impact of detention on young adults or to monitor indicators of social network destruction. These types of insights and methods can be used both for advocating against detention as well as to help identify specially dramatic detentions.
\\
\item \textbf{Combat xenophobia and discrimination.} 
A growing body of research within computational social science has been devoted to untangling complex societal issues, from polarization~\cite{bail2018exposure}, community integration~\cite{lamanna2018immigrant}, gender and ethnic stereotypes~\cite{garg2018word}, to fake news~\cite{guess2019less,bovet2019influence}. 
Xenophobia and discrimination are equivalent issues and deserve equal attention.
Further, while a majority of studies focus on detection of signals and quantification of discriminatory practices, little work has been devoted to developing intervention strategies for addressing these issues~\cite{valente2012network}.
\\
\item \textbf{Protect refugee and migrant children from exploitation and violence.} 
Social networks are critical for information access and for general safety. Past research has already demonstrated how network analysis can be applied to design more efficient interventions to for example reduce conflict in schools~\cite{paluck2016changing}. Mobile phone records have also shown great potential in untangling these complex issue and have been used to study individual communication capacities~\cite{miritello2013limited}, behavioural adaptation~\cite{eagle2009community}, and detection of unusual behaviors~\cite{dobra2015spatiotemporal}. As such, these methodologies have the potential to radically improve the toolbox of child protection systems -- if applied within proper ethical and governance frameworks. 
\end{itemize}

\section{Discussion}

In 2017 the world experienced record breaking displacement numbers. Every day 44,400 new people were forcibly displaced from their homes; more than a four-fold increase since 2003. Today around 68.5 million people are forcibly displaced, a number greater than the population of the UK, and children have been estimated to account for up to 52\% of the total number of displaced individuals~\cite{unhcr2018_2}. The sheer size of this problem, along with its growing complexity urgently requires that we strengthen our efforts and look for new ways to improve response, preparedness, data and understanding.

The existing body of research on big data, network analysis and complex systems science has shown promise in providing fresh and powerful new perspectives and tools to curve this issue. There are, nevertheless, some key challenges that need to be solved for it to happen.
At a scientific level, it is of key importance that these disciplines include vulnerable populations at the core of their analysis and efforts. Much of the research conducted so far, including advancements in computational social sciences, has been done looking at data-rich populations in high-income countries. As such, a majority of these findings and methodologies might not generalize to vulnerable populations, especially children. Therefore, special attention has to be drawn on validating relevant findings for this unique context, keeping in mind the representativeness of the data, and ensuring that the most vulnerable populations are the focus of new research efforts.
Initiatives as the D4R challenge are key to drive and convene scientists into these critical challenges. 
Nevertheless, additional efforts have to be undertaken to ensure this ecosystem of innovation and research continues on a daily basis. 

Access to data (e.g. mobile phone data) is one of they key challenges organizations such as UNICEF face in order to drive and incorporate data-driven methods into operations. 

Privacy, ethics and transparency are also key concerns to have in mind while working on these issues.
To build equitable technologies we need to include the most vulnerable populations from day one into our scientific methodologies and organizational frameworks, otherwise we run the risk of building a more unequal society.
However, this cannot be done naively. It is paramount to ensure that envisioned solutions or derived insights cannot be used to discriminate.



Official data on refugees, what in the scientific community is considered ground truth, has its limitations too. While a child is a child, political frameworks differentiate between refugees and migrants, this can ultimately leave displaced children out of some official statistics. Thus, relying blindly on official numbers might replicate some of these shortfalls and weaknesses into new methodologies.
Collaborations and a deeper understanding of the humanitarian and development ecosystems can empower the broader scientific community to understand the limitations of official statistics and even look beyond them to identify populations suffering similar conditions despite different legal definitions.  


To succeed, it is important to embrace diverse data sharing frameworks for both research and operations~\cite{de2018privacy}; to build inclusive platforms that integrate findings and tools into active response systems~\cite{magicbox}, and to reframe scientific questions such that they include the most vulnerable contexts by strengthening collaborations between scientists and humanitarian communities.




\bibliographystyle{spphys}
\bibliography{references}
\end{document}